\documentclass[epj]{svjour}

\usepackage{widetext}

\usepackage{amssymb}
\usepackage{amsfonts}
\usepackage{bm}
\usepackage{epsfig}
\usepackage{epsf}
\usepackage[]{graphicx}

\usepackage{color}
\usepackage{cite}
\usepackage{xspace}

\renewcommand{\d}{{\rm d}}
\newcommand {\ee}{{\rm e}}
\renewcommand {\i}{{\rm i}}

\newcommand {\bfn} {{\bf n}}
\newcommand {\bfp} {{\bf p}}
\newcommand {\bfr} {{\bf r}}
\newcommand {\bfv} {{\bf v}}

\newcommand {\bfE} {{\bf E}}

\newcommand {\bfR} {{\bf R}}

\newcommand {\calA} {{\cal A}}

\newcommand {\E}  {{\varepsilon}}
\newcommand {\om}  {{\omega}}
\newcommand {\Om}  {{\Omega}}

\newcommand{\Lch}{{L_{\rm ch}}}

\begin{document}


\title
{Radiation Emission by Electrons Channeling in Bent 
Silicon Crystals}

\author{
R G Polozkov\inst{1},
V K Ivanov\inst{1},
G B Sushko\inst{2},
A V Korol\inst{2,3},
and 
A V Solov'yov\inst{2}}
%
\institute{St. Petersburg State Polytechnical University,
Politechnicheskaya 29, 195251 St. Petersburg, Russia \and  Department of Physics, Goethe University, 
Max-von-Laue-Str. 1, 60438 Frankfurt am Main, Germany \and St. Petersburg State Maritime University, 
Leninsky ave. 101, 198262 St. Petersburg, Russia}
%

%
\date{Received: date / Revised version: date}
%
\abstract{
Results of numerical simulations of electron  channeling
and emission spectra are reported for straight and uniformly bent  silicon crystal. 
The projectile trajectories are computed using the newly developed module
\cite{NewPaper_2013} of the MBN  Explorer package 
\cite{MBN_ExplorerPaper,MBN_ExplorerSite}.
The electron channeling along Si(110)  
crystallographic planes is studied for the 
projectile energy 855 MeV.
}

\authorrunning{Polozkov et al.} 
\maketitle
\section{Introduction}

The basic effect of the channeling process in a straight crystal 
is in an anomalously large distance which a charged projectile can penetrate 
moving along a crystallographic plane (the planar channeling) or an axis 
(the axial channeling) 
experiencing collective action of the electrostatic field of the 
lattice ions \cite{Lindhard}.
The field is repulsive for positively charged projectiles so that 
they are steered into the inter-atomic region.
Negatively charged particles, electrons in particular, move in the close vicinity 
of ion strings.

\textcolor{black}{
The channeling can also occur in a bent crystal provided the 
bending radius $R$ is large enough in comparison with the critical one
$R_{\mathrm{c}}$ 
\cite{Tsyganov1976}.
}

Channeling of a charged particle is accompanied by the channeling radiation 
\cite{ChRad:Kumakhov1976} which is due to channeling oscillations of the 
particle under the action of an interplanar or an axial field.
The radiation intensity depends on the type of the projectile and on its energy
as well as on the type of a crystal and a crystallographic plane (axis).
\textcolor{black}{
The phenomenon of channeling radiation of a charged projectile in a straight crystal is well known
(see, for example,  Refs.
\cite{Instrum,Andersen_ChanRadReview_1983,BakEtal1985,BakEtal1988,
BazylevZhevago,RelCha,Kumakhov2,SaenzUberall1981,Uggerhoj1993,Uggerhoj_RPM2005,
Uggerhoj2011}).
}
\textcolor{black}{
Various aspects of radiation formed in crystals bent with the constant curvature 
radius $R$ were discussed in 
Refs.
\cite{ArutyunovEtAl_NP_1991,Bent_Rad:Arutyunov1,Bashmakov1981,KaplinVorobev1978,
SolovyovSchaeferGreiner1996,Taratin_Review_1998,TaratinVorobiev1988,TaratinVorobiev1989}
although with various degree of detail.}

\textcolor{black}{
The motion of a channeling particle in a bent crystal contains two components:
the channeling motion and the translation along the centerline of the bent channel.
The latter motion gives rise to the synchrotron-type radiation \cite{Landau2,Jackson}.  
Therefore, the total spectrum of radiation formed by an {ultra-relativistic} 
projectile in a bent crystal contains the features of channeling radiation 
and those of synchrotron radiation.
}
\textcolor{black}{
The condition of stable channeling in a bent crystal, $R \gg R_{\mathrm{c}}$, 
\cite{Tsyganov1976,BiryukovChesnokovKotovBook} implies
that the bending radius $R$ exceeds greatly the (typical) curvature radius of 
the channeling oscillations. 
Therefore, the synchrotron radiation modifies mainly the soft-photon part of the 
emission spectrum.
}
\textcolor{black}{
Study
} 
\textcolor{black}{
of  this part of the spectrum is especially important 
in connection with the concept  of a crystalline undulator
\cite{KSG1998,KSG_review_1999}.
}
By means of crystalline undulator it is feasible to produce monochromatic
 undulator-like radiation 
in the hundreds of keV up to the MeV photon energy range. 
The intensity and characteristic frequencies of the radiation can 
be varied by changing the type of channeling particles,
the beam energy, the crystal type  and the parameters of periodic bending
(see recent review \cite{ChannelingBook2013}  for more details).

In recent years, several experiments were carried out to measure the 
channeling parameters and the characteristics of emitted radiation 
of sub-GeV light projectiles.
These include the attempts made  \cite{BaranovEtAl_CU_2006} 
or planned to be made \cite{Backe_EtAl_2011a,Backe_EtAl_2008} 
to detect the radiation from a positron-based crystalline undulator.
More recently, a series of the experiments with straight, bent  and 
periodically bent crystals
have been carried out with 195--855 MeV electron beams  
at the Mainz Microtron (Germany) facility
\cite{Backe_EtAl_2010,Backe_EtAl_2011,Backe_EtAl_2013}.
The crystalline undulators, used in the experiment, 
were manufactured in Aarhus University (Denmark) using the molecular beam
epitaxy technology to produce strained-layer Si$_{1-x}$Ge$_{x}$ 
superlattices with varying germanium content 
\cite{MikkelsenUggerhoj2000,Darmstadt01}.
\textcolor{black}{
Another set of experiments with diamond crystalline undulators is planned within the 
E-212 collaboration at the SLAC facility (USA) with 10\dots 20 GeV electron 
beam \cite{SLAC_FACET}.
}

Theoretical support of the ongoing and future experiments as well as accumulation
of numerical data on channeling and radiative processes of ultra-relativistic 
projectiles in crystals of various content and structure must be based on an 
accurate procedure which allows one to simulate the trajectories 
corresponding to the channeling and non-channeling regimes.
Recently, a universal code to simulate trajectories
of various projectiles in an arbitrary scattering medium, either structured 
(straight, bent and
periodically crystals, superlattices, nanotubes etc) or amorphous (solids, 
liquids) has been developed as a new module
\cite{NewPaper_2013} of the MBN  Explorer package 
\cite{MBN_ExplorerPaper,MBN_ExplorerSite}.
To simulate propagation of particles through media the algorithms used in modern 
molecular dynamics codes were utilized.
Verification of the code against available experimental data
as well as against predictions of other theoretical models 
were carried out for electron and positron channeling in straight Si(110) 
as well as in amorphous Si \cite{NewPaper_2013}.
\textcolor{black}{
In more recent papers \cite{BentSilicon_2013,Sub_GeV_2013} the simulations were
extended to the case of bent and periodically bent Si(110) and Si(111) channels.
}
In the cited papers, critical analysis was carried out of the underlying physical model 
and the algorithm implemented in the recent code for electron channeling described in 
Refs. \cite{KKSG_simulation_straight}.
It was shown, that the specific model for electron--atom scattering
leads in a noticeable overestimation of the mean scattering angle
\textcolor{black}{
 and, as a result,
to the underestimation of the dechanneling length.
}%

In this paper we present new results on electron channeling
and emission spectra in straight and uniformly bent silicon crystal. 
The electron channeling along Si(110) 
crystallographic planes are studied for the 
projectile energy  $\E$ = 855 MeV for two lengths of a crystalline sample and 
for different bending curvatures.

\section{Electron Channeling in Si (110) \label{ElectronChanneling}}

\subsection{Simulation of the Channeling Process with MBN Explorer
\label{MBN_algorithm}}

\textcolor{black}{
To perform 3D simulation of the propagation of an ultra-relativistic 
projectile through a crystalline medium two additional features 
were added to the molecular dynamics algorithms used in the \textsc{MBN Explorer} 
package \cite{MBN_ExplorerPaper}.
The first feature concerns the implementation and integration of the 
relativistic equations of motion.
The second one is the dynamic generation of the crystalline medium.
In detail, these features are described in \cite{NewPaper_2013}.
Below in this section we outline the key points only.
}

\textcolor{black}{
Within the framework of classical mechanics 
the motion of an ultra-relativistic projectile of the charge $q$ and
mass $m$ in an external electrostatic field $\bfE(\bfr)$
is described by the equations:
\begin{eqnarray}
\partial \bfp /\partial t = q\bfE,
\qquad
\partial \bfr /\partial t = \bfv.
\label{MBN_algorithm:eq.01} 
\end{eqnarray}
}
\textcolor{black}{
Here $\bfp = m\gamma\bfv$ is momentum,
$\gamma = [1-(v/c)^2]^{-1/2} = \E/(mc^2) \gg 1$ is the Lorenz factor, $\E$ is 
the projectile energy and $c$ is the speed of light. 
Equations (\ref{MBN_algorithm:eq.01}) are to be integrated for $t\geq 0$ using the
initial values of the coordinates $\bfr_0=\bfr(0)$ and velocity
$\bfv_{0}=\bfv(0)$ of the particle at the crystal entrance.
}

\textcolor{black}{
To describe the motion in a scattering medium (a crystal, in particular) 
it is important to compute accurately and efficiently
the electrostatic field due to the medium atoms.
In the channeling module of \textsc{MBN Explorer} the field  is calculated as
$\bfE(\bfr) = -\partial U(\bfr) /\partial \bfr$,
where electrostatic potential $U(\bfr)$ is a sum of atomic 
potentials
\begin{eqnarray}
U(\bfr)
=
\sum_{j} U_{\mathrm{at}}\left(\rho_j\right)\Bigl|_{\boldsymbol{\rho}_j=\bfr - \bfR_j },
\label{MBN_algorithm:eq.02} 
\end{eqnarray}
where $\bfR_j$ stands for the position vector of the $j$-th atom.
The atomic potentials $U_{\mathrm{at}}$ are evaluated within the Moli\`{e}re approximation,
see, e.g.,  \cite{Baier}.
}

\textcolor{black}{
Formally, the sum in Eq.~(\ref{MBN_algorithm:eq.02}) is carried out over all 
atoms of the crystal.
However, accounting for a rapid decrease of $U_{\mathrm{at}}\left(\rho_j\right)$ 
at the distances $\rho_j\gg a_{\mathrm{TF}}$ from the nucleus
(here the Thomas-Fermi radius $a_{\mathrm{TF}}$ is chosen to estimate the mean atomic 
radius),
one can introduce the cutoff $\rho_{\max}$ above which the
contribution of $U_{\mathrm{at}}\left(\rho_j\right)$ is negligible.
Therefore, for a given observation point $\bfr$ the sum can be restricted
to the atoms located inside a sphere of the radius $\rho_{\max}$.
To facilitate the search for such atoms the linked cell algorithm 
is employed.
As a result, the total number of computational operations is reduced 
considerably.
}

\textcolor{black}{
To simulate the channeling motion along a particular crystallographic
plane with Miller indices $(klm)$ the following algorithm is 
used \cite{NewPaper_2013}.\footnote{Axial channeling can also be simulated. 
For the sake of clarity, here we refer to the case of planar channeling.}  
}

\textcolor{black}{
As a first step, a crystalline lattice is generated inside the simulation box
of the size $L_x\times L_y \times L_z$.
The $z$-axis is oriented along the beam direction and is parallel to
the $(klm)$ plane, the $y$ axis is perpendicular to the plane.
To avoid the axial channeling (when not desired) the $z$-axis
is chosen to be not collinear with major crystallographic axes.
The position vectors of the nodes $\bfR_j^{(0)}$ ($j=1,2,\dots, N$) 
within the simulation box are generated in accordance with the 
type of the Bravais cell of the crystal and using the pre-defined values of the 
the lattice vectors.
}

\textcolor{black}{
Once the nodes are defined, the position vectors of the 
atomic nuclei are generated with account for random displacement 
$\boldsymbol{\Delta}_j$ from the nodal positions
due to thermal vibrations.
The Cartesian components $\Delta_{j\alpha}$, $\alpha=x,y,z$,  
are normally distributed:
\begin{eqnarray}
w(\Delta_{j\alpha})
=
\frac{1}{\sqrt{2\pi u_T^2}} \exp\left(- {\Delta_{j\alpha}^2 \over 2u_T^2}\right)\,.
\label{MBN_algorithm:eq.03} 
\end{eqnarray}
Here $u_T$ is the root-mean-square amplitude of thermal vibrations.
Its values for various crystals at room temperature can be found in 
\cite{Gemmel}.
}

\textcolor{black}{
Integration of equation (\ref{MBN_algorithm:eq.01})
starts at $t=0$ when the particle ``enters'' the crystal at $z=0$.
The initial coordinates $x_0$ and $y_0$ are randomly chosen
to be lying in the central part of the $(xy)$-plane of the sizes $\Delta x = 2d$,
$\Delta y = d$ where $d$ is the interplanar spacing of the $(klm)$ planes.
The initial velocity $\bfv_0=(v_{0x},v_{0y},v_{0z})$ is predominantly oriented 
along $z$, i.e. the conditions $v_{0z}\approx c \gg v_{0x},v_{0y}$ are implied.
The transverse components $v_{0x}, v_{0y}$ can be chosen with account for the 
beam emittance.
}

\textcolor{black}{
To simulate the propagation of a particle through a crystal of finite thickness
$L$ a new type of boundary conditions, the ``dynamic simulation box'', has been 
implemented in \textsc{MBN Explorer} \cite{NewPaper_2013}.
This algorithm implies the following.
}

\textcolor{black}{
A projectile moves within the simulation box interacting with the atoms lying
inside the cutoff sphere.
Once the distance $l$ from the projectile to the nearest face becomes 
$l\approx \rho_{\max}$ a new simulation box of the same size is generated
with its geometrical center coinciding (approximately) with the
position of the projectile.
To avoid spurious change in the force $q\bfE$,
the positions of the atoms located in the intersection of the old and the new
simulation boxes are not changed.
In the rest part of the new box the positions of
atomic nuclei are generated following the scheme described above.
}

\textcolor{black}{
The motion in the amorphous medium can also be simulated.
For doing this it is necessary to avoid incidental alignment of the initial
velocity $\bfv_0$ with major crystallographic directions.
This regime is useful to calculate the spectral and spectral-angular
distribution of the incoherent bremsstrahlung.
}

\textcolor{black}{
To simulate a bent crystal, the coordinates $(x',y',z')$ of each lattice node 
are obtained from the coordinates 
$(x,y,z)$ of the same node in the straight crystal according to the relations  
$x' = x$, $y' = y + \delta y(z)$, $z' = z$, where
\begin{equation}
\delta y(z) = R - \sqrt{R^2-z^2} \approx {z^2 \over 2R} 
\label{MBN_algorithm:eq.04} 
\end{equation}
is the shape of the bent crystallographic plane with bending radius $R$.
The approximate equation is valid if $z\leq L \ll R$.
}

\textcolor{black}{
In a bent crystal, the channeling condition \cite{Tsyganov1976} implies that 
the centrifugal force $F_{\rm cf}= pv/R \approx \E/R$ is smaller than the maximum 
interplanar force $F_{\max}$.
Thus, the value of the bending parameter 
\begin{equation}
C = {F_{\rm cf} \over F_{\max}} = {\E \over R F_{\max}},
\label{MBN_algorithm:eq.05} 
\end{equation}
must be much less than one.
The case $C = 0$ characterizes to the straight crystal.
The value $C = 1$ corresponds to the critical (minimum) 
radius $R_{\rm c}=\E/F_{\max}$
at which the potential barrier between the channels disappear. 
The value $F_{\rm max}$ can been estimated withing the 
continuous approximation for the planar potential.
For Si(110) one can use $F_{\rm max} = 5.7$ GeV/cm at room temperature 
\cite{BiryukovChesnokovKotovBook}.
}

\subsection{Numerical Results for $\E=855$ MeV Electron Channeling in 
Si(110)
\label{ChannelingResults}}

\textcolor{black}{
MBN Explorer was used to simulate the trajectories 
855 Mev electrons in straight and bent silicon crystals incident 
at $z=0$ along the (110) crystallographic planes.
The calculations were performed for two values of the crystal length,
$L=$ 25 and 75 $\mu$m, measured along the beam direction.
}

\begin{figure} [h]
\centering
\includegraphics[scale=0.35,clip]{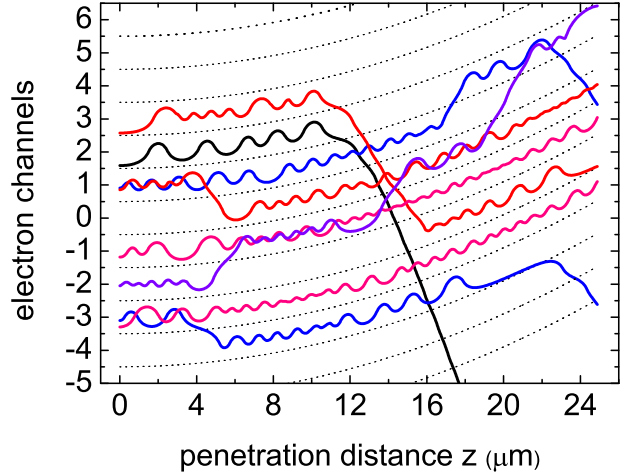}  
\caption{
Channeling of 855 MeV electrons in a  $L_1=25$ $\mu$m thick silicon crystal
with bending radius $R = 1.3$ cm (the bending parameter $C \approx 0.1$). 
}
\label{Figure.01} 
\end{figure}

\textcolor{black}{
Several selected trajectories, which refer to the crystal bent with radius 
$R=1.3$ cm, are presented in Fig. \ref{Figure.01}.
They illustrate a variety of features which characterize the electron
motion: the channeling mode, the dechanneling, the over-barrier motion,
the rechanneling process, rare events of hard collisions etc.
First, let us note that since the dechanneling length of a 855 MeV electron  
in straight Si(110) is $\approx12$ $\mu$m 
\cite{NewPaper_2013,BentSilicon_2013}, 
it is not surprising that the events of channeling through the 
whole crystal are quite rare.
For the indicated value of the bending radius only $\lesssim 1$ \% of the incident
particles travel through $L=$ 25 $\mu$m thick crystal moving in the channeling mode
in the same channel where they were captured at the entrance. 
On the other hand, the events of rechanneling, i.e., capture to the channeling 
mode of an over-barrier particle, are quite common for electrons 
\cite{KKSG_simulation_straight,NewPaper_2013,BentSilicon_2013}.
Even the multiple rechanneling events are not rare.
The rechanneling process, as well as the dechanneling one, is due to the random 
scattering of a projectile from the crystal constituents.
In each scattering event the projectile can either increase or decrease its transverse
energy $\varepsilon_y$.
The sequence of the events with $\Delta\varepsilon_y>0$ can lead to the 
dechanneling of initially channeled particle.
The opposite process, resulting in a noticeable reduction of $\varepsilon_y$
of the over-barrier particle, can also occur leading to the capture of the
particle into the channeling mode, i.e., the rechanneling.
}

\textcolor{black}{
The simulated trajectories were used to estimate the dechanneling length 
(in the case of the electron channeling) and to calculate spectral distribution
of the emitted radiation.
}

\begin{table*}
\caption{Acceptance $\calA$, penetration lengths $L_{p1}$, $L_{p2}$,  and 
total channeling length $\Lch$ for  $\E=855$ MeV electrons channeled 
in straight ($R=\infty$) and bent ($R<\infty$) Si(110) crystals.
The data refer to the crystal length $L=75$ $\mu\mbox{m}$. 
$N_0$ stands for the number of simulated trajectories.
Bending parameter $C$ is calculated from (\ref{MBN_algorithm:eq.05}) using
the value $F_{\max}=5.7$ GeV/cm.}
\label{Table_lengths}
\begin{tabular*}{1.00\textwidth}%
{@{\extracolsep{\fill}}ccccccc}
\hline\noalign{\smallskip}
   $R$  & $C$    &$\calA$ &  $L_{p1}$    &  $L_{p2}$   &  $\Lch$          & $N_0$ \\
   (cm) &        &        &  ($\mu$m)    &  ($\mu$m)   &  ($\mu$m)        & \\
\noalign{\smallskip}\hline\noalign{\smallskip}
$\infty$&  0.0   &   0.66 &$11.72\pm 0.32$&$10.88\pm 0.15$&$26.24\pm 0.49$  &  12600 \\ 
70      &  0.002 &   0.65 &$11.71\pm 0.33$&$10.86\pm 0.16$&$25.78\pm 0.50$  &  11950 \\ 
25      &  0.006 &   0.64 &$11.62\pm 0.33$&$10.86\pm 0.16$&$24.59\pm 0.50$  &  11950 \\ 
13      &  0.01  &   0.65 &$11.43\pm 0.32$&$10.87\pm 0.17$&$21.79\pm 0.50$  &  12000 \\ 
7       &  0.02  &   0.64 &$11.55\pm 0.34$&$11.06\pm 0.21$&$16.35\pm 0.55$  &  12000 \\ 
2.5     &  0.06  &   0.55 &$10.23\pm 0.30$&$10.16\pm 0.27$&$07.00\pm0.27$  &  12000 \\ 
1.3     &  0.1   &   0.44 &$08.67\pm0.28$&$08.71\pm0.27$&$04.14\pm0.21$  &  12050  \\ 
0.75    &  0.2   &   0.34 &$07.03\pm0.41$&$07.03\pm0.40$&$02.41\pm0.27$  &   4300 \\ 
0.45    &  0.3   &   0.22 &$05.32\pm0.33$&$05.33\pm0.33$&$01.18\pm0.17$  &   4500   \\ 
\noalign{\smallskip}\hline
\end{tabular*}
\end{table*}

\textcolor{black}{
Although the particles are initially directed along the (110) planes, not all of 
them become captured into the channeling mode at the crystal entrance.
The important parameter to estimate is the acceptance,
which is defined as a ratio of the number $N_{\mathrm{acc}}$ 
of particles captures into the channeling mode at the entrance (the accepted particles)
to the total number $N_{0}$ of the simulated trajectories 
(the incident particles):
\begin{equation}
{\cal A} = {N_{\mathrm{acc}} \over N_0}\,.
\label{ChannelingResults:eq.01}  
\end{equation}
}
\textcolor{black}{
To quantify the dechanneling process of the accepted particles the 
following two {\em penetration} depths $L_{\mathrm{p}}$ were introduced in 
Ref.~\cite{NewPaper_2013}.
The first one, notated below as $L_{\mathrm p1}$, is found as a mean value
of the primary channeling segments, which started at the entrance and lasted till
the dechanneling point somewhere inside the crystal.
Generally speaking, this quantity is dependent on the angular distribution
of the particles at the entrance.
The second penetration depth, $L_{\mathrm p2}$, is defined as  a mean value 
of all channeling segments, including those which are due to the rechanneling.
In the rechanneling process an electron is captured into the channeling mode
having, statistically, an arbitrary value of the incident angle $\Theta$
not greater than  Lindhard's critical $\Theta_{\mathrm L}$ angle \cite{Lindhard}.
Therefore, $L_{\mathrm p2}$ mimics the penetration depth of the beam with
a non-zero emittance $\approx \Theta_{\mathrm L}$.\footnote{This statement 
implies that crystal thickness $L$ is large enough to ensure $L \gg L_{\mathrm p2}$.}
}

\textcolor{black}{
Either one from $L_{\mathrm p1}$ and $L_{\mathrm p2}$ provides an estimate of 
the {\em dechanneling length}. 
The results of calculations of these quantities based on the simulations of 
855 MeV electron channeling in straight ($R=\infty$) and bent ($R<\infty$) Si(110)  
are presented in Table~\ref{Table_lengths}.
In addition to $L_{\rm p1,2}$ the total channeling length $\Lch$, 
defined as an average length of all channeling segments per trajectory,
was calculated.
Also included in the table are the number $N_0$ of simulated 
trajectories,  acceptance $\cal A$ and bending parameter $C$.
The critical radius in Si(110) for an $\E=855$ MeV projectile is 
 $R_{\rm c}=\E/F_{\max} = 0.15$ cm.  
}

\textcolor{black}{
The calculated values of $L_{\mathrm p1}$ and $L_{\mathrm p2}$ 
exceed noticeably the dechanneling lengths estimated for $\E=855$ MeV electrons 
channeled in straight \cite{KKSG_simulation_straight} and bent 
 \cite{Kostyuk_EPJD_2013} Si(110) channels. 
The discrepancy, being on the level of 30  per cent for the straight channel, 
becomes more pronounced for $C>0$ reaching 100 per cent for $C=0.5$ 
(not indicated in the table). 
Most probable, the discrepancy is due to a peculiar model used in 
\cite{KKSG_simulation_straight,Kostyuk_EPJD_2013} to describe electron--atom 
elastic scattering.
The model substitutes the atom with its ``snapshot'' image: instead of the 
continuously distributed electron charge the atomic electrons are treated 
as point-like charges placed at fixed positions around the nucleus.
The interaction of an ultra-relativistic projectile with each atomic constituent 
is treated in terms of the classical Rutherford scattering with zero recoil 
for the scatterer.
In Ref. \cite{NewPaper_2013} it was demonstrated, 
that the ``snapshot'' model noticeably overestimates the mean scattering angle
in a single electron-atom collision.
The mean square angle for a single scattering is an important quantity
in the multiple-scattering region, where there is a large succession of small-angle 
deflections symmetrically distributed about the incident direction.
In particular, the mean square  angle due to soft collisions defines the diffusion 
coefficient which, in turn, is proportional to the dechanneling length 
(see, for example, Refs. \cite{BiryukovChesnokovKotovBook,Backe_EtAl_2008}).
}

\section{Radiation Emitted by Electrons in Si (110) 
\label{Radiation}}

The simulated trajectories in straight and periodically bent
Si(110) channels were used to compute
spectral distribution of the emitted radiation.
To this end, for each set of simulated trajectories of the total number 
$N_{0}$ the spectral distribution emitted within the cone 
$\theta\leq\theta_0$ with respect to the incident beam
was calculated as follows:
\begin{eqnarray}
{\d E(\theta\leq\theta_0) \over \hbar\, \d \om}
=
{1 \over N_0}
\sum_{j=1}^{N_0} 
\int\limits_{0}^{2\pi}
\int\limits_{0}^{\theta_0}
{\d^3 E_j \over \hbar\,\d \om\, \d\Om}
\label{Radiation:eq.01} 
\end{eqnarray}
Here, 
$\d^3 E_j/\hbar\d\om\, \d\Om$ 
stands for the spectral-angular distribution 
emitted by a particle which moves along the $j$th trajectory.
The sum is carries out over {\em all} simulated trajectories, 
i.e. its takes into account the contribution of the channeling 
segments of the trajectories as well as of those corresponding to the
non-channeling regime.

To calculate $\d^3 E_j/\hbar\d\om\, \d\Om$  one can use
a general quasi-classical method developed by Baier and Katkov 
\cite{Baier67}.
The details of the formalism, as well as  its application to a variety
of radiative processes, can be found in Ref.~\cite{Baier}
(see also \cite{Uggerhoj2011}).
A remarkable feature of this method is that it allows one to combine
the classical description of the motion in an external field
and the quantum effect of radiative recoil, i.e.
the change of the projectile energy due to the  photon emission.
Its role is governed by the ratio $\hbar \om /\E$.
In the limit  $\hbar \omega /\varepsilon \ll 1$ a purely classical
description of the radiative process can be used (see, e.g., \cite{Landau2,Jackson}).
For  $\hbar \omega /\varepsilon \le 1$ quantum corrections must be accounted for.
The quasi-classical approach explicitly takes into account the quantum corrections 
due to the radiative recoil.
The method is applicable in the whole range of the
emitted photon energies, except for the extreme high-energy tail of the spectrum
$\left(1-\hbar\om/\E\right)\ll 1$.

Within the framework of quasi-classical approach 
the spectral distribution of energy radiated in given direction $\bfn$ by an 
ultra-relativistic particle
is given by the the following expression (see Ref. \cite{Baier} for the details):

\begin{widetext}
\begin{equation} \label{Radiation:eq.02} 
{\d^3 E \over \hbar\d\om\, \d \Om}
=
\alpha \,
{ q^2\omega^2  \over 8\pi^2 }
\int\limits_{-\infty}^{\infty} \d t_1\!
\int\limits_{-\infty}^{\infty} \d t_2\,
\ee^{\i \,\omega^{\prime} \left(\psi(t_1) -\psi(t_2)\right)}
\left[
\left( 1+(1+u)^2 \right)
\left(
{\bfv_1\cdot\bfv_2 \over c^2}  -1
\right)
+{u^2 \over \gamma^2}
\right],
\end{equation}
\end{widetext}
\noindent where $\alpha= e^2/ \hbar\, c$ is the fine structure constant,
$q$ is the charge of a projectile in units of the elementary
charge, $\bfv_{1,2} =\bfv(t_{1,2})$ denote the velocities, and
the phase function reads as $\psi(t) = t - \bfn\cdot\bfr(t)/ c$.
The quantities $\om^{\prime}$ and $u$ account for the radiative recoil:
\begin{eqnarray}
\om^{\prime}
=
(1+u)\, \om,
\qquad
u =
{\hbar \om \over \E - \hbar \om}.
\label{Radiation:eq.03} 
\end{eqnarray}
In the classical limit
$u\approx \hbar\om /\E\to 0$ and $\om^{\prime} \to \om$, so that
(\ref{Radiation:eq.02}) reduces to the classical formula \cite{Landau2,Jackson}.

Eq. (\ref{Radiation:eq.02}) allows us to compute the emission spectrum 
for each simulated trajectory. 

The simulated trajectories were used to compute
spectral distribution of the emitted radiation following the 
algorithm described in Refs.~\cite{NewPaper_2013}. 
The calculations were carried out for 25 and 75 $\mu$m thick straight and bent crystals.
The integration over the emission angle in (\ref{Radiation:eq.01}) was performed
for the aperture $\theta_{0}$ = 2.4 mrad which greatly exceeds
the natural emission angle  $\gamma^{-1} \approx 0.6$ mrad.
The averaging was carried out over $N_0\approx 2000$ simulated trajectories.

\subsection{Emission Spectra for $L=25$ $\mu$m}

\begin{figure} [h]
\centering
\includegraphics[scale=0.35,clip]{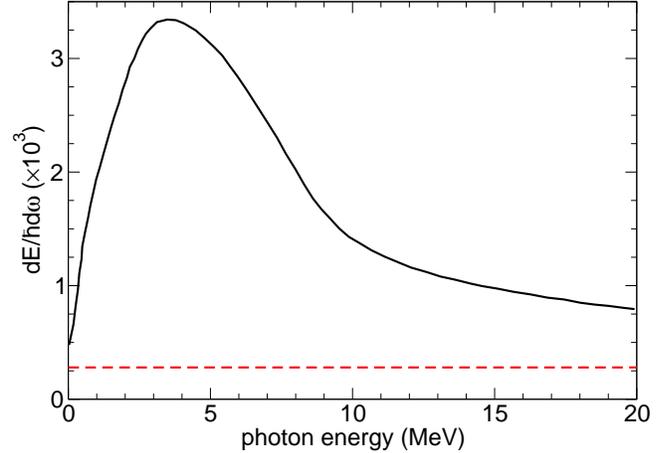}  
\caption{
Solid curve represents the calculated spectral distribution of radiation by 
855 MeV {electrons} in oriented straight Si(110).
Dashed line shows the Bethe-Heitler for amorphous silicon.
The data refer to the aperture $\theta_{0}$ = 2.4 mrad and to the crystal 
thickness $L=25$ $\mu$m.
}
\label{Figure.02} 
\end{figure}

In Fig.~\ref{Figure.02} the calculated spectral distribution 
(\ref{Radiation:eq.01})
of the channeling radiation in straight Si(110) crystal (solid curve) 
is compared with the spectral intensity in an amorphous silicon (dashed curve) 
obtained within the Bethe-Heitler (BH) approximation (see, for example,
Ref. \cite{Tsai1974}).   
It is seen that the intensity of radiation in the oriented crystal greatly exceeds
(in the maximum located at $\hbar\om \approx 4$ MeV there is an order of magnitude excess)
that in the amorphous medium. 
The enhancement is due to the emission by the particles moving along quasi-periodic 
channeling trajectories, which bear close resemblance with the undulating motion. 
As a result, constructive interference of the waves
emitted from different but similar parts of the trajectory increases the intensity. 

\begin{figure} [h]
\centering
\includegraphics[scale=0.35,clip]{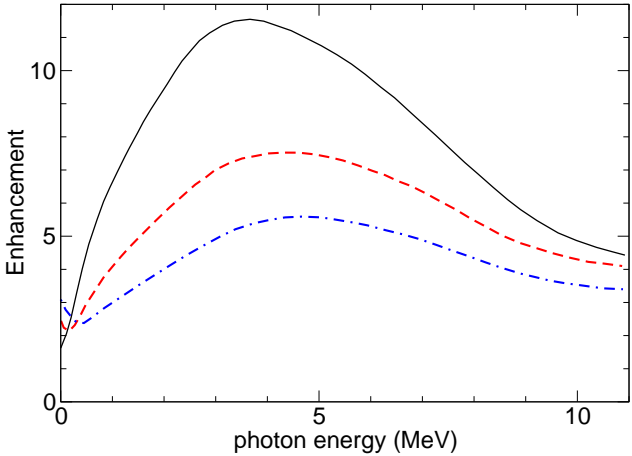}  
\caption{
Enhancement factor for 855 MeV {electron} channeled in 
$L=25$ $\mu$m straight (solid curve) and bent Si crystals along (110) planes.
Broken curve corresponds to the bending radius $R=2.5$ cm (the bending parameter 
$C=0.06$), 
chained curve -- to $R=1.3$ cm ($C=0.1$). 
The data refer to the aperture $\theta_{0}$ = 2.4 mrad.
}
\label{Figure.03} 
\end{figure}

Spectral enhancement factor, i.e. the spectral distribution $\d E/\d (\hbar\om) $
of radiation emitted in the crystalline medium normalized to that in amorphous 
silicon, calculated for 855 MeV electrons is presented in Fig.~\ref{Figure.03}.
Solid curve represents the dependence obtained for straight Si(110),
two other curves   stand for uniformly bent channel:
the broken curve corresponds to $R=2.5$ cm, the chained one -- to 
$R=1.3$ cm.
  
Let us note several specific features of the emission spectra formed in bent channels.

First, the bending gives rise to the synchrotron radiation, since the channeled particle
experiences the circular motion in addition to the channeling oscillations.
This leads to the increase of the intensity in the photon energy range $\lesssim 10^2$ 
keV. 
For these energies the radiation yield from the bent channel
exceeds that from the straight channel.
%

Second, there is a noticeable decrease in the intensity of the channeling radiation
with the decrease of the bending radius.
The ratio $\xi$ of the maximum value of the spectrum in the bent channel to 
that in the straight one is $0.66$ for $R=2.5$ cm and $0.49$ for $R=1.3$ cm.
The statistical uncertainties of these values due to the finite number of the 
simulated trajectories is on the level of 10 per cent.
The decrease of $\xi$ can be explain as follows.
In the vicinity of the maximum, the main contribution to the channeling radiation
intensity comes from those segments of a trajectory where the particle stays in the 
same channel experiencing channeling oscillations.
The intensity of radiation emitted from either of this segments and integrated over 
the emission angles is proportional to the segment length $l$ and to the square of the 
undulator parameter $K$ associated with the channeling oscillations.
The undulator parameter $K$ is related to the mean square of the transverse velocity:
$K^2 = 2\gamma^2 \,\langle v^2_{\perp}\rangle /c^2$ (see, e.g., \cite{Baier}).
Then, to estimate the right-hand side of Eq. (\ref{Radiation:eq.01}) one can
consider the contribution to the sum coming from the primary channeling
segments.
Hence
\begin{eqnarray}
{\d E(\theta\leq\theta_0) \over \hbar\, \d \om}
\propto 
\calA\, K^2\, L_{\mathrm p1}\,.
\label{Radiation:eq.04} 
\end{eqnarray} 
For bending radii much larger than the critical radius 
$R_{\rm c}=\E/F_{\max} = 0.15$ the factor $K^2$ depends weakly on $R$.
Hence, the factor $\xi$, defined above, can be calculated at the ratio
of the products $\calA\, L_{\mathrm p1}$ calculated for the bent and the
straight channels.
Using the data from Table \ref{Table_lengths} one estimates:
$\xi = 0.72 \pm 0.03$ and $0.48 \pm 0.02$ for $R=2.5$ and 1.3 cm, respectively.  
These values correlate with the ones quoted above.
Thus, the decrease in the acceptance $\calA$ and in the penetration length
$L_{\mathrm p1}$ with $R$ is the main reason for lowering the intensity of 
channeling radiation.

Finally, comparing the curves in Fig.~\ref{Figure.03} one notices that
the position $\om_{\max}$ of the maximum shifts to higher photon energies 
with the growth of the crystal curvature $1/R$.
This feature is specific for the electron channeling (more generally, for 
channeling of negatively charged projectiles) and is due to strong anharmonicity
of the channeling oscillations.
As a result, the frequency $\Om_{\rm ch}$ of the oscillations varies with
the amplitude $a_{\rm ch}$.
Typically, smaller vales of $a_{\rm ch}$ correspond to larger frequencies
(see, for example, the results of simulations presented in \cite{Sub_GeV_2013}).
As the curvature increases, the allowed values of $a_{\rm ch}$ decrease
due to the action of the centrifugal force \cite{BiryukovChesnokovKotovBook}.
Hence, on average, the frequency of the channeling increases and so does 
the frequency of the emitted photons $\om_{\max} \propto \Om_{\rm ch}$. 

The spectra presented in Figs.~\ref{Figure.02} and \ref{Figure.03} 
are obtained from Eq. (\ref{Radiation:eq.01}) by averaging  
over all simulated trajectories. 
The main contribution to the radiation enhancement comes from the electrons 
which stay mostly in the channeling regime. 
In this context it is of interest to analyze the spectral distribution of radiation 
produced only by electrons propagating through the whole bent crystal staying 
in a single channel or in a few different channels (i.e. changing the channels
due to the rechanneling effects).  
Such trajectories, although being quite rare,  allow one to visualize the
influence of the channeling oscillations as well as of the 
rechanneling effect on the synchrotron-like part of the spectrum.

\begin{figure} [h]
\centering
\includegraphics[scale=0.35,clip]{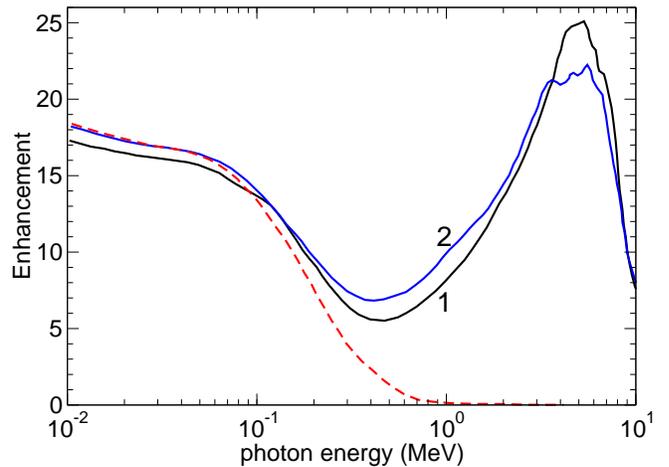}  
\caption{
Spectral distribution of radiation by 855 MeV {electrons} in a $L=25$ $\mu$m 
thick Si(110) crystal bend with the radius $R=1.3$ cm.
The data refer to the aperture $\theta_{0}$ = 2.4 mrad.
Curves 1 and 2 correspond to the enhancement factor averaged over two 
specially selected sets of trajectories.
The broken represents the spectrum of synchrotron radiation normalized to the 
Bethe-Heitler background.
See also explanation in the text.
}
\label{Figure.04}
\end{figure}

\textcolor{black}{
To carry out this program we analyzed $N_0=2000$ electron trajectories simulated in
$L=25$ $\mu$m Si(110) crystal bent with the radius $R=1.3$ cm.
Fifteen trajectories ("group 1") were found corresponding to the motion through the whole 
crystal staying in the same channel in which the projectile was captured at the 
entrance. 
Another twelve trajectories ("group 2") were comprised of several channeling 
segments of the total length $L$.
Examples of these trajectories (from the both groups) one finds in 
Fig.~\ref{Figure.01}.
The radiation spectra, averaged over each group of electrons, are represented in  
Fig.~\ref{Figure.04} by solid curves labeled with "1" and "2".
At $\hbar\om \approx 5$ MeV both curves have the maxima corresponding 
to the channeling radiation.
It is not surprising that the enhancement factor 
averaged over the specially selected trajectories is much larger than the one
averaged over all trajectories, see broken curve in Fig.~\ref{Figure.03}.  
}

The increase of the enhancement factor at small photon energies is associated 
with the synchrotron radiation.  
To visualize this we calculated the spectral distribution of radiation due to motion 
along a $25$ $\mu m$ arc of a circle of the radius 1.3 cm, i.e. along the bent 
channel centerline ignoring the channeling oscillations.
The result, normalized to the Bethe-Heitler spectrum,  
is presented in Fig.~\ref{Figure.04} by the broken line.
Within the statistical errors (not indicated) the curves 1 and 2 
coincide with the synchrotron spectrum at $\hbar\om\lesssim 10^{-1}$ MeV.  

\begin{figure} [h]
\centering
\includegraphics[scale=0.35,clip]{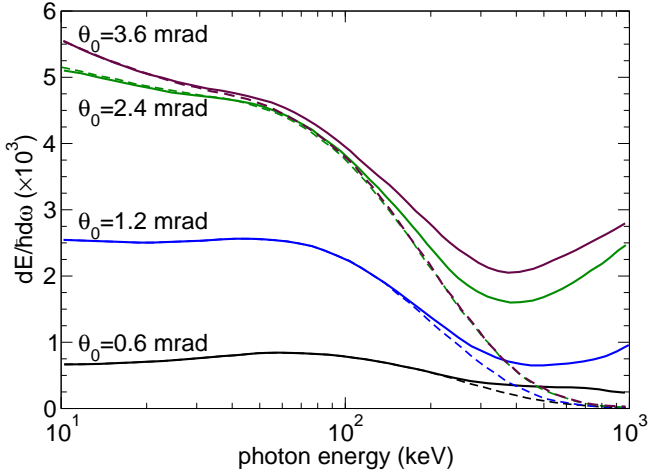}  
\caption{
Low-energy part of radiation spectrum formed by 855 MeV electrons in a  
$L=25$ $\mu$m thick silicon crystal with bending radius $R = 1.3$ cm.
The curves correspond to different apertures, which are integer multiples of 
the natural emission angle $\gamma^{-1}\approx 0.6$ mrad.
Full curves stand for the spectra averaged over the electrons channeling
through whole crystal in the same channel, 
Broken curves represents the spectra of synchrotron radiation.
}
\label{Figure.05}
\end{figure}

In Ref.~\cite{Sub_GeV_2013} the influence of the detector aperture $\theta_0$ 
on the form of the spectral distribution of the channeling radiation in straight
and bent Si(110) channels was explored.
In Fig.~\ref{Figure.05} we present results of the similar analysis of the 
low-energy part of the spectrum.
Full curves correspond to the spectra (\ref{Radiation:eq.01}) averaged over the 
simulated trajectories 
 corresponding to the motion through the whole crystal staying in the same channel 
("group 1").
Broken curves represent the distribution of the synchrotron radiation due
to the motion along the circle arc.
The calculations were performed for $\theta_0$ = 0.6, 1.2, 2.4 and 3.6 mrad
which are integer multiples of  the natural emission angle 
$\theta_{\gamma} = \gamma^{-1}\approx 0.6$ mrad. 
The latter characterizes the cone (along the instantaneous velocity)
which accumulates most of the radiation emitted by an ultra-relativistic
projectile.

Two features of the presented dependences can be mentioned.

First, it is seen that the intensity is quite sensitive to 
the detector aperture.
This effect is more pronounced for the lower values of $\theta_0$ and 
within the $10\dots 100$ keV photon energy range where most of radiation is
emitted via the synchrotron mechanism.
For example, a two-fold change in the aperture from $\theta_{\gamma}$
to $2\theta_{\gamma}$ results in a nearly four-fold increase of the intensity 
at in the lowest-energy part of the spectrum. 
Such a behaviour can be understood if one compares the  quoted values of $\theta_0$
with the angle of rotation of the bent crystal centerline, 
$\theta_{L} = L/R \approx 1.9$ mrad.
For apertures smaller than $\theta_{L}$ the radiation within the 
cone $\theta_0$ along the incident velocity will be effectively emitted 
only from the initial part of the crystal, the length $l_0$ of which can be estimated
as $l_0 \approx R\theta_0 = (\theta_0/\theta_{L})L < L$.    
Thus, the raise of the aperture from $0.6$ to $1.2$ mrad results in the two-fold
increase of $l_0$ which, in turn, leads to additional augmentation of the emitted 
intensity.
This effects becomes less pronounced for larger apertures, 
$\theta_0 > \theta_{L} \gg \theta_{\gamma}$, which collects virtually all radiation
emitted within $L$.

Another feature to be mentioned is that the influence of the channeling motion on 
the emitted spectrum becomes more pronounced over wider range of photon energies 
with the increase of the aperture. 
Indeed, the deviation of the total emission spectrum 
(full curves) from the spectrum of synchrotron radiation (broken curves) starts 
at $\hbar\om\approx 250$ keV for $\theta_0=0.6$ mrad but
at $\hbar\om\approx 50$ keV for $\theta_0=3.6$ mrad.
To explain this one recalls that channeling motion bears close resemblance with the 
undulating motion. 
As a result, constructive interference of the waves
emitted from different but similar parts of the trajectory increases the intensity. 
For each value of the emission angle $\theta$ the
coherence effect is most pronounced for the radiation into harmonics.
The frequency  of the lowest (first) harmonic can be estimated as
$\om_{1}(\theta) = 2\gamma^2\, \Om_{\rm ch}/\left(1 + \gamma^2 \theta^2 + K^2/2\right)$
(see, e.g., \cite{Baier}).
Anharmonicity of the electron channeling oscillations results in the variation
of $\Om_{\rm ch}$ along the trajectory. 
Therefore, the radiation will be emitted within some frequency band which will
form the main peak in the spectral distribution of the channeling radiation
(see Fig. \ref{Figure.02}). 
For small apertures, when $(\gamma\theta_0)^2 \ll 1$, the emission of 
low-energy photons with $\om \ll \om_{1}(0)$ is strongly suppressed.
For larger apertures a big part of energy is radiated into the cone
$\gamma^{-1} < \theta < \theta_{0}$. 
For $(\gamma\theta_{0})^2\gg 1$ the harmonic energy is strongly red-shifted
$\om_{1}(\theta_0) \ll \om_{1}(0)$.
As a result, the contribution of the channeling radiation to the low-energy part of the
spectrum increases with the aperture.

\subsection{Emission Spectra for $L=75$ $\mu$m}

\begin{figure} [h]
\centering
\includegraphics[scale=0.35,clip]{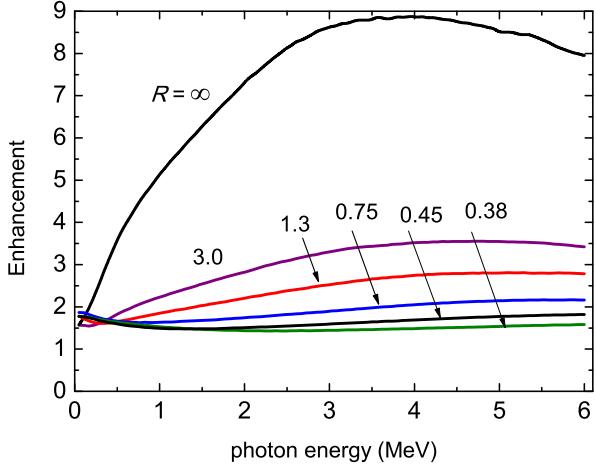}  
\caption{
Enhancement factors
for 855 MeV {electrons} channeled in 
$L=75$ $\mu$m straight and bent Si(110) crystals.
The numbers indicate the values of the bending radius in cm.
The data refer to the aperture $\theta_{0}$ =2.4 mrad, 
}
\label{Figure.06}
\end{figure}

\begin{figure} [h]
\centering
\includegraphics[scale=0.35,clip]{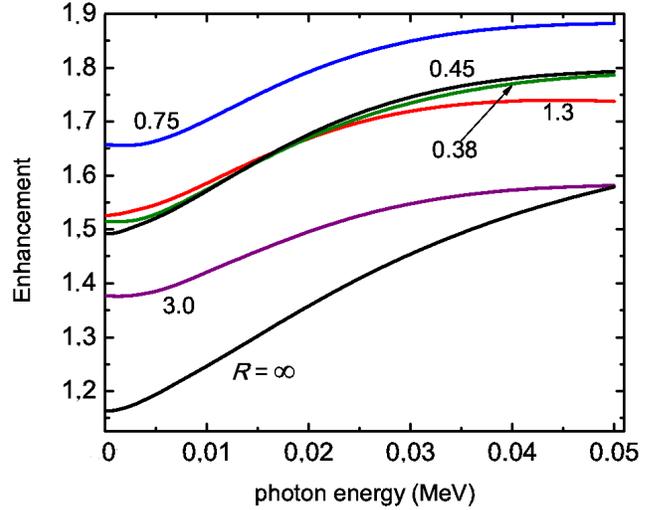}  
\caption{
Enhancement factors for the low-energy part of 
radiation spectrum formed by 855 MeV electrons in a $L=75$ $\mu$m straight and
bent Si(110).
The numbers indicate the values of the bending radius $R$ in cm.
The data refer to the aperture $\theta_{0}$ =2.4 mrad, 
}
\label{Figure.07}
\end{figure}

\textcolor{black}{
For $L=75$ $\mu$m thick Si(110) crystal  
the trajectories were simulated for a wider range of the bending radius 
(in addition to the values indicated in Table \ref{Table_lengths} the calculations 
were performed for $R=0.3$ cm, see Ref. \cite{BentSilicon_2013}).
The results of calculations of the enhancement factor obtained for the 
aperture $\theta_{0} = 2.4$ mrad  are presented 
in Figs. \ref{Figure.06} and \ref{Figure.07}.
}

Figure \ref{Figure.06} illustrates the modification of the spectral dependence of 
the enhancement factor in the vicinity of the maximum of channeling radiation.
Comparing the presented dependences with those calculated for a shorter crystal, 
see Fig. \ref{Figure.03},
one can state that in both cases the profiles of the enhancement factor are similar,
 and the maximum values of the enhancement factor are close in both cases.
In both figures there are two curves calculated for the same values of 
the bending radius: $R=\infty$ (straight crystal) and $R=1.3$ cm.
Comparing these one notices that the maximum values are larger 
(by factors $\approx 1.4$ and $1.6$, respectively)
for a shorter crystal.
To explain this one can use the following arguments.
In a straight crystal, the intensity of channeling radiation is proportional to the 
total length of the channeling segments, $L_{\rm ch}$, whereas the intensity of 
the incoherent bremsstrahlung scales with the crystal length $L$.
Hence, the enhancement factor is proportional to $L_{\rm ch}/L$.
For a $75$ $\mu$m thick crystal $L_{\rm ch}\approx 26$ $\mu$m 
(see Table \ref{Table_lengths}).
For $L=25$ $\mu$m we found $L_{\rm ch}\approx 12.6$ $\mu$m.
Hence the ratio of the $L_{\rm ch}/L$ values calculated for  $L=25$ and $75$ $\mu$m
straight Si(110) crystals is $3/2$ which correlates with the factor quoted above.
Similar arguments can be applied to the case of a bent crystal.
The only difference is that the intensity of the background bremsstrahlung radiation
integrated over the aperture $\theta_0$ is proportional to the effective length 
$l_0 \sim \min\left\{L, \theta_0 R\right\}$.


The synchrotron radiation influences the total spectrum in the photon
energy range well below the maximum of channeling radiation. 
Fig. \ref{Figure.07} illustrates the behaviour 
the enhancement factor in the low-energy part of the spectrum.

It is seen from the figure that the enhancement due to the 
synchrotron radiation is a non-monotonous function of the bending radius.
At small curvatures the enhancement increases with $1/R$. 
The maximum values are achieved at $R\approx 0.75$ cm and then the enhancement 
decreases with the curvature.
This feature is due to the finite aperture  which for 
sufficiently small values of $R$ introduces the length 
$l_0 < L$ of the crystal where the radiation detected within the cone
$\theta \leq \theta_0$ is effectively formed, and thus reduces the enhancement.


\section{\textcolor{black}{Conclusion}}

Using a newly developed code \cite{NewPaper_2013}, 
which was implemented as a module in the MBN Explorer package 
\cite{MBN_ExplorerPaper},
we have performed numerical simulations of trajectories of  
ultra-relativistic electrons in oriented straight and bent 
single Si(110) crystals.

The simulated trajectories were used as the input data for numerical analysis 
of the intensity of the emitted radiation.
In the case of straight crystals the channeling radiation appears 
atop the incoherent bremsstrahlung background. 
In a bent channel the spectrum is enriched by the synchrotron radiation due to the
circular motion of the projectile along the bent centerline.

The calculation of the spectra as well as the numerical analysis of channeling 
conditions and properties (acceptance, dechanneling length) 
have been carried out for the beam of  855 MeV electrons for two lengths of Si crystal. 
The obtained and presented results are of interest in connection with  
the ongoing experiments with electron beams at Mainz Microtron \cite{Backe_EtAl_2011}.

\section{Acknowledgments}
We are grateful to Hartmut Backe and Werner Lauth for fruitful and
stimulating discussions. 
The work was supported by the European Commission CUTE-IRSES project 
(grant GA-2010-269131). 
The possibility of performing complex computer 
simulations at the Frankfurt Center for Scientific Computing is 
gratefully acknowledged.


\end{document}